\documentclass[12pt,preprint]{aastex}

\begin{document}
\title{Interactions between a Bright YSO and the MSX Infrared-Dark Cloud
G79.3+0.3: An Early Stage of Triggered Star Formation?}
\author {R. O. Redman\altaffilmark{1}, P. A. Feldman\altaffilmark{1},
F. Wyrowski\altaffilmark{2},
S. C\^ot\'e\altaffilmark{1},
S. J. Carey\altaffilmark{3},
M. P. Egan\altaffilmark{4}}
\altaffiltext{1}{NRC of Canada, 5071 W. Saanich Road, Victoria, BC V9E 2E7, 
Canada; Russell.Redman@nrc.ca, Paul.Feldman@nrc.ca, Stephanie.Cote@nrc.ca}
\altaffiltext{2}{Astronomy Program, University of Maryland, College Park, MD 
20742, USA. Current address: Max-Planck-Institut-f\"{u}r-Radioastronomie, Auf
den H\"{u}gel 69, Bonn, D-53125, Germany; wyrowski@mpifr.bonn.mpg.de}
\altaffiltext{3}{Institute for Scientific Research, Boston College, 140 
Commonwealth Ave., Chestnut Hill, MA 02467, USA. Current address: IPAC, Caltech,
770 South Wilson Avenue, Pasadena, CA 91125, USA; carey@ipac.caltech.edu}
\altaffiltext{4}{Air Force Research Laboratory, VSBC, 29 Randolph Rd.,
Hanscom AFB, MA 01731-3010, USA. Current Address: MDA/AT,
7100 Defense, Pentagon, Washington, DC 20301-7100, USA; Michael.Egan@bmdo.osd.mil}

\begin{abstract} 
Millimeter and mid-infrared observations have been made of the dense clumps of
dust and gas and of young stellar objects (YSOs)  associated with the bright,
compact submillimeter source G79.3+0.3~P1 in the  relatively nearby MSX
infrared-dark cloud  G79.3+0.3. The Gemini mid-infrared observations reported
here indicate the presence of three YSOs within the cloud.  BIMA 3~mm continuum
observations show that the brightest  of the YSOs is likely to be a Herbig
Ae/Be star.  High-angular-resolution  molecular-line observations suggest that
a wind from this star may be triggering collapse in the adjacent molecular
cloud.  The submillimeter source G79.3+0.3~P1 itself does not contain infrared 
sources and may represent an earlier stage of star formation. 
\end{abstract}


\section{Introduction} \citet{ega98} have identified a large population of cold
dust clouds in images of the Galactic Plane made with the MSX satellite
\citep{pri01}.  These clouds appear as dark patches of  absorbing
material against a background of mid-infrared emission bands and emission from 
warm, small dust grains.  Egan et
al. defined the infrared-dark clouds (IRDCs) as having one to several 
magnitudes of extinction at 8~\micron\  and no obvious emission in any of the 8
to 25~\micron\ bands observed by the MSX satellite.  They concluded that the
IRDCs possess hundreds of magnitudes of visual extinction and contain large
column densities of cold dust. Observations of a selection of these clouds in
H$_2$CO by \citet{car98} revealed that much of the gas inside the IRDCs has
T$_{\rm K}$~$\approx$~10-20~K and n(H$_2$)~$\gtrsim$~10$^6$~cm$^{-3}$, with
H$_2$ column densities ranging up to 10$^{23}$~cm$^{-2}$.  Most of these
objects are quite distant, with kinematic distances between 1 and 8.5 kpc.  

The IRDCs appear to be a very interesting population of molecular clouds. On
theoretical grounds, we expect that molecular clouds on the verge of  star
formation will contain extremely cold and dense condensations, which will show
up as IRDCs if they are favourably located in front of a region of extended  
mid-IR emission.  The selection criteria for the MSX IRDCs \citep{car98} 
ensures that  they do not yet contain high-mass main-sequence stars and HII
regions.  

Trying to relate the physical properties of this selection of dust clouds to
those of better-studied nearby molecular clouds, such as those in the 
Orion-Taurus region, is problematic.  Most of  the best studied clouds lie in
the outer Galaxy where there is insufficient  background emission for IRDCs to
be identifiable.  By contrast, the large distances of many MSX IRDCs make them
difficult to detect at visible  and near-infrared wavelengths, and difficult to
pick out from the surrounding  clutter of lower density clouds in
molecular-line surveys.  

A notable exception is the cloud G79.3+0.3, located in the Cygnus Rift  at an
estimated distance of 800~pc \citep{mil37, ikh61, wen91}, close enough to the
Sun that it can be seen at visible wavelengths \citep{red00b}. Figure~\ref{g79}
and Figure~\ref{8mic}
show our JCMT SCUBA 850~\micron\ image of the G79.3+0.3 IRDC and the MSX
8~\micron\ image of the same region, respectively. Unlike the other
IRDCs we have observed, this cloud exhibits a variety of indicators of star
formation.  Most prominently, the HII region DR15
(G79.307+0.277) lies behind and slightly south
of the IRDC, between P4 and P5 in the figure.  It is not
immediately obvious whether the IRDC is part of a larger complex including 
DR15 or is an unrelated foreground cloud. In  visible and near-IR images of the
region \citep{red00b}, there is an association of stars (to the west of DR15) 
whose northern boundary is defined by the southern edge of the IRDC.  If DR15
is related to this  association, it suggests that both DR15 and the association
lie beyond the IRDC.  In the 8~\micron\ MSX image (Figure~\ref{8mic}), the dark
filament of  the IRDC (whose dust content is traced by the SCUBA 850~\micron\
emission in Figure~\ref{g79}) is  interrupted between P1 and P3 by a bright patch of warm dust
emission which is suggestive of deeply embedded hot stars. Because of this, the
eastern and western parts of the IRDC were initially assigned different
designations in the MSX catalog 
(G79.34+0.33 and G79.27+0.38, labeled G79.34 and G79.27 in Figure~\ref{g79}), 
although the figure clearly shows them to comprise a single, connected
filament. A Herbig-Haro jet that appears to be driven by a YSO at
RA(2000) = 20~31~45.5, Dec(2000) = +40~18~44 (see
Figure~\ref{8mic}) has been discovered in front of the IRDC at visible
wavelengths \citep{red00b}. 

In this paper we focus our attention on the star-forming activity in the
vicinity of  its most prominent condensation, G79.3+0.3~P1. This region was
selected for closer study because comparison of the SCUBA and MSX images
discussed below showed the presence of a faint, point-like emission source
MSX5C G079.3398+00.3415 in the MSX image that was nearly coincident with the
brightest compact source of emission at 850 and 450~\micron . This suggested
the presence of a deeply embedded star interacting with the cold dust cloud,
but the relatively coarse resolution of both the MSX and SCUBA images hampered
a more detailed interpretation. To  address these issues, we combined  JCMT
observations at 450~\micron , BIMA interferometry at 3~mm,  and Gemini North
imaging at 10.75~\micron , together with archival data from the MSX satellite 
and the 2MASS survey.

\section{Observations}

\subsection{JCMT Observations}

JCMT SCUBA observations were obtained at 850 and 450~\micron\ where   the
resolution is $14\arcsec$ and $8\arcsec$,  respectively
\citep{hol99}.  ``Scan mapping'' was employed to produce  Nyquist sampled
images. Sky  subtraction was provided by chopping to a nearby  location. We
used the Emerson-2 scan-mapping procedure 
\citep{eme95} as adapted for
SCUBA \citep{jen98a} with offsets of $20\arcsec$,
$30\arcsec$, and $65\arcsec$ at position angles of
$0\arcdeg$  and $90\arcdeg$. The  data reduction process for scan maps attempts
to correct for the chopping,  so  that sources appear only once, as positive
features.  However, structures  larger than the chop throw may not be
reproduced correctly. The zero-level  is uncertain unless the maps are large
enough to include some blank sky in  the image. Descriptions of SCUBA and its
observing modes may  be found  in \citet{hol99} and \citet{jen00}.

The SCUBA observations were taken under fair weather conditions  
($\tau_{850}$= 0.24-0.30) on 1999 April. The data were
processed using the standard routines provided by the SURF package
\citep{jen98b} for flat fielding, extinction correction,
and sky subtraction. Uranus were used as a flux-density calibrator,
using the calculated flux densities of Uranus produced by the FLUXES 
program, which makes use of  the
measurements reported by \citet{gri93}.  Our measured  flux 
densities are good to $\pm 15$\%. Pointing was performed
on Uranus and G34.3, and should be better than $2\arcsec$. 

Observations of the HCO$^+(3-2)$ transition were made using RxA3 in a
raster-scanning mode.  The grid of sample points had a spacing of $10\arcsec
\times 10\arcsec$, centered at RA = 20~32~03.4, Dec=+40~19~20 in J2000
coordinates and oriented with the Y-axis of the grid at a position angle
$-94\fdg 4$ to follow the overall orientation of the G79.3+0.3 IRDC.  The
nominal beam size of the JCMT at this wavelength is 18\farcs 4, so the grid of
samples is approximately Nyquist spaced.  With the raster-mapping technique,
the beam is smeared somewhat in the direction of motion of the telescope during
the sample, giving an effective beam size of $20\farcs 5\times 18\farcs 4$,
with the long axis aligned north-south.  In this paper we report on a subset 
of the whole dataset, a square region $160\arcsec\times 160\arcsec$ with a
position angle for the vertical axis of $0\arcdeg$ centered at RA=20~32~22.24,
Dec=+40~19~53 (J2000), covering the same area as the BIMA HCO$^+(1-0)$
observations discussed below.  

The HCO$^+(3-2)$ observations were taken on the night of 2001 June 26 (UT)
under weather conditions which are fairly poor for the JCMT but still quite
usable at 267~GHz. The zenith opacity at 225~GHz 
ranged from 0.14 to 0.20 and the seeing was greater than 1\arcsec . The
Starlink program SPECX was used to process the observations into a datacube.

\subsection{BIMA Observations}

The Berkeley-Illinois-Maryland-Association interferometer
\citep{wel96}, a.k.a. the BIMA array, was used in its B, C, and D
configurations to observe G79.3+0.3~P1 at 3~mm (90~GHz) for a total of
5 tracks between 1999 September and 2000 May. The lower and upper sideband
signals were combined to obtain a total continuum bandwidth of 0.9~GHz.
Additionally, two tracks in C configuration were obtained in 1999 September at
245~GHz by observing a 7-point mosaic to cover the 3~mm primary beam. 
The correlator setup included the HCO$^+~(1-0)$ line at
89.188~GHz and the CS$~(5-4)$ line at 244.935~GHz.  MWC~349 and
J2025+337 were observed as phase calibrators every 15 to 25
minutes. MWC~349 (assuming fluxes of 1 and 1.75 Jy at 90 and 245 GHz,
respectively) was used to check the flux-density calibration derived
from observations on Uranus. The errors in the flux-density scale are
$\le 20\%$ (30\%) at 3~mm (1.2~mm).  Calibration and image
deconvolution was carried out using the MIRIAD package \citep{sau95}.  
The 3mm continuum and CS$~(5-4)$ images were deconvolved using a
clean algorithm (MIRIAD task ``clean'' and ``mossdi''), while the
HCO$^+~(1-0)$ image was deconvolved with the MIRIAD task
``maxen''. Beam sizes of $5\farcs 3\times 5\farcs 0$, 
$4\farcs 5\times 4\farcs 2$, and $3\farcs 3\times 2\farcs 7$ were
obtained for the images at 3~mm continuum, HCO$^+~(1-0)$ and CS$~(5-4)$,
respectively.  The rms
noise in the 3~mm continuum image and the line averaged HCO$^+~(1-0)$ and
CS$~(5-4)$ emission images are 0.8, 56 and 500~mJy/beam, respectively.

\subsection{Gemini Observations}

Observations were carried out at the Gemini North Telescope on 2000  December 7
and 2001 July 10 using OSCIR, the U. Florida mid-IR  imager/spectrometer.
OSCIR  employs Si:As blocked-impurity-band (BIB) detector array with 128$\times
$128 pixels, optimized for a wavelength coverage from 8 to 25 \micron . On
Gemini its plate scale is  0\farcs 089 per pixel, which gives a FOV of 
11\farcs 4 $\times$11\farcs 4. All observations  were made using the
standard chop/nod technique to remove the sky and  telescope background
emission, and employing a chop frequency of 3~Hz and a throw  of 25\arcsec .
Images were obtained in the N-band ($\lambda _0 = 10.75~\micron$, $\Delta
\lambda = 5.23~\micron$), with 60 seconds of on-source integration time per
field. A 3$\times$3 mosaic was obtained by  combining five fields acquired in
December (mainly the SE part of the mosaic),  with six  acquired in July.  

The standard stars $\beta$~Peg and Vega  were used to provide flux
calibration.  Their flux densities were taken to  be 352.13~Jy and 37.77~Jy, 
respectively,  through the OSCIR N-band filter  \citep{fis01}. Our 2001 July
run was plagued  by light cirrus and higher humidity; hence the flux of
2MASSI~2032220+402017,  observed the previous December, was used to estimate
the fluxes of the fainter sources detected in the July fields. 

All data were taken with the telescope active optics in open-loop,  using an
elevation-dependent lookup table, and with fast guiding provided by  the
secondary mirror ``tip-tilt'' corrections under control of the  Peripheral Wave
Front Sensor (PWFS) locked on a nearby guide star. Hence some residual
astigmatism was present in the point-spread function. The detected  sources
should be  considered unresolved since their measured FWHMs (at best
$0\farcs 47$; the diffraction limit in N-band for Gemini is $0\farcs 28$)  are
similar to  those  measured for the imaged  standard stars. There is a
discrepancy of order 2\arcsec\ in the coordinates of the detected  source 
between the December data and the July  data. The adopted coordinates
are those derived in July, since a guide star  with better position accuracy
was used (error $<$~0\farcs 2).  However, the accuracy of the model of the
PWFS arm 2 for the distance between the OSCIR center and the PWFS is unknown. 
There is a 1\farcs 5 offset between the position of 2MASSI~2032220+402017,
with a nominal accuracy of 0\farcs 2, and the apparent position of the OSCIR
source.  We do not believe this difference is real because of the possible
systematic error mentioned above, so in this paper we assume that these two
sources are the same.

\section{Results}
\subsection{YSOs and Infrared Stars}
 
The original motivation for the N-band Gemini observations reported here was to
determine whether the MSX 8~\micron\ source was composed of a single star,  a
cluster of fainter stars, or a diffuse emission patch, and to determine what
relationship if any it had to the P1 source in the SCUBA images.  In fact, 
most of the mid-IR emission comes from a single point-like source
coincident with the star 
2MASSI~2032220+402017 (see Figure~\ref{p1}a).  

In addition, two fainter sources are also found in the field, but far enough
from the central source that they are unlikely to be directly associated with
it.  Figure~\ref{p1}b shows the locations of the three stars superimposed on  a
contour map of the SCUBA 450~\micron\ emission.  The angular resolution of the
450~\micron\ map is 7\arcsec , sufficient to demonstrate that
2MASSI~2032220+402017 lies a full beamwidth away from the peak of the
450~\micron\ emission.  In fact, none of the three N-band sources lie within
the top contours of the 450~\micron\ emission from the P1 source.  If YSOs are
forming within the  dust clumps constituting the P1 source, they are too faint
and/or too deeply  embedded to be detectable in these images. 

From the N-band images it is impossible to rule out the presence of a
significant component of extended emission around the 2MASS star.  As mentioned
above, the PSF had a FWHM of $0\farcs 47$, with roughly 60\% of the signal
contained in a circle of diameter $0\farcs 8$ centered on the star. An
additional 40\% of the signal comes from a region extending out as far as
$2\farcs 7$ from the star.  Because we were unable to take observations of a
reliable PSF calibration star,  it is possible that as much as 40\% of the
emission could be extended on a scale larger than $0\farcs8$.  Alternatively,
the apparent extended emission could simply be due to the wings of the PSF.

The photometric properties of all three N-band sources are summarized in
Table~\ref{3stars}.  The two fainter sources have been calibrated relative to
2MASSI~2032220+402017.  In Table~\ref{3stars}, the uncertainty in the 
flux-density estimates has been divided into internal errors (which indicate
the significance of the detection) and systematic errors due largely to the
unknown degree of possible extended emission around the 2MASS star.  The flux
density of 2MASSI~2032220+402017 was determined using a large aperture to
capture all  of the signal.  The photometry of the weaker sources were measured
in two ways.  Firstly, assuming the apparent extended emission is due to the
wings of the PSF, the photometry was done by convolving the images with a PSF
derived from the bright source, effectively using a small aperture to maximize
the signal-to-noise ratio.  Secondly, assuming that the apparent emission is
genuinely extended, the photometry was done using a $2\farcs7$ aperture
comparable to that used to calibrate the 2MASS star.  The reported fluxes in
Table~\ref{3stars} are the average of the two measured values.

On probabilistic grounds we can determine that the two weak N-band sources are
YSOs, most likely associated with the IRDC, and are not simply background 
field stars seen through the cloud.  The (N-K$_S$) colors of unreddened stars 
are quite small, so we can estimate the surface density of N-band stars by
using the 2MASS survey to  measure the brightness distribution at K$_S$ in
nearby fields with low extinction. The two N-band stars have N magnitudes of
7.5 and 8.4.  If the stars were merely background stars, the corresponding K
magnitudes would be similar, hundreds of times brighter than we would expect to
find in the small area of sky we observed. 

It is interesting that there appears to be a population of heavily reddened
stars visible in the 2MASS K$_S$-band image of G79.3+0.3~P1 (which sometimes are
not even visible at H or J). We conclude that all three N-band sources are part
of a much larger cluster of embedded and highly reddened young stars that has
formed in the general area  around G79.3+0.3~P1.  

With the available data we can only place a lower limit on the reddening toward
2MASSI~2032220+402017.  It is normal for the near- and mid-infrared emission from 
YSOs to be dominated by their circumstellar disks.  The intrinsic H-K colors of
these objects are larger than those of stars [e.g., see \citet{mal98}], but are
still small compared to the apparent reddening of 2MASSI~2032220+402017.  Since
the 2MASS star is close to the most heavily extincted part of the IRDC, we will
assume that most of the reddening is interstellar. 2MASSI~2032220+402017 has an
apparent  brightness $K_S$ from the 2MASS Point Source Catalog of $K_S=12.0$
but is not detected at H-band ($H> 17.3$). From these measurements we can
estimate $A_K > 8~{\rm mag}$ using the extinction law from \citet{mar90},
assuming that the intrinsic H-K color of 2MASSI~2032220+402017 is small.   With
this estimate for $A_K$, the unreddened apparent magnitude  $K_S < 4~{\rm
mag}$. 

At an assumed distance of 800~pc, the absolute magnitude of 
2MASSI~2032220+402017 at $K_S$ would be brighter than -5.5~mag.  Although this
ostensibly is the absolute magnitude of an  early-O star,  the absence of a
prominent H~II region indicates that the star must be considerably fainter and
cooler with a large IR excess.   This object must be an extremely luminous
pre-main sequence star with the energy output of an early-B star. It thus seems
likely that 2MASSI~2032220+402017 is a Herbig Ae/Be star with a large IR
excess  \citep{mal98} due a  circumstellar disk and/or dust in a strong ionized
wind.  

As discussed by \citet{fue02}, pre-main-sequence stars that will ultimately
have spectral types earlier than B5 pass through a stage lasting several
hundred thousand years during which they are surrounded by a massive
circumstellar disk but are still too cool to produce a significant UV flux.  
In this stage, a stellar wind from the pre-main-sequence star will be the
primary mechanism dispersing the surrounding cloud. 

\subsection{Continuum Emission Sources Containing Dust and Hot Gas}

To investigate further the nature of 2MASSI~2032220+402017, BIMA observations
with an angular resolution of  5\arcsec\ were made of the 3~mm continuum
emission.  At this resolution, the source breaks up into three components as
shown in Figure~\ref{p1}c. The observed properties of the three components
measured from the BIMA 3~mm image are given in Table~\ref{fluxden}. Two
components, labeled A and B, accord well with the morphology of our  SCUBA
450~\micron\ map.  Their measured flux densities are consistent with the
emission arising from dust, as discussed below.   The third 3~mm component,
labeled C, coincides within the accuracy of the astrometry with
2MASSI~2032220+402017, but is not evident in the 450~\micron\ SCUBA image. It
must have a significantly  smaller spectral index, consistent with the 3~mm
emission being dominated by  thermal bremsstrahlung from the ionized wind of a
Herbig Ae/Be star  \citep{ski93}.

We estimated the dust temperatures and total masses of components A and B 
following \citet{hil83}.  Assuming that the dust opacity per unit total mass 
column density (dust and gas) varies as $\kappa_\nu = \kappa_0 (\nu /\nu_0)^\beta$,
the dust temperatures were estimated as functions of $\beta$ from the ratios of
the flux densities at 450~\micron\ and 3~mm.  The masses were calculated
using 
\begin{equation}
\label{massform}
M_{tot} = D^2 F_\nu / \kappa_\nu B_\nu(T)
\end{equation}
where $M_{tot}$ is the total mass, $D$ is the distance, and
$B_\nu(T)$ is the Planck function for a dust temperature $T$.   We adopted the
value of $\kappa_0 = 0.005$~cm$^2$~gm$^{-1}$ at  1.3~mm from \citet{mot98},
and references therein, as an appropriate value for pre-stellar dense cores and
clumps.  Table~\ref{masstab} gives the derived dust temperatures and total
masses as functions of $\beta$ for the two sources.

If the flux densities for components A and B had been measured separately at a
third  frequency, it would be possible to estimate $\beta$ for each component. 
Unfortunately, the beam size of the JCMT at 850~\micron\ is so large that
components A, B and C all blend together, and the contribution to the combined
flux density from component C is unknown.  Nevertheless, we can still derive
bounds on the allowed range of $\beta$. The estimated temperatures in both
components become unphysically large as $\beta$ drops below 1.5. 
Also, if we assume that  $\beta$ is the same in
components A and B, we can  invert Equation~\ref{massform} to predict their
combined flux density at 850~\micron\ as a function of $\beta$.  For $\beta >
2.0$, the predicted value of the combined flux density becomes greater than
2.0~Jy, the
observed flux density of  the whole source at 850~\micron . Thus, $\beta$ is bounded in the
range $1.5 \le \beta \le 2.0$.  

Other observers have measured values for $\beta$ in similar clouds with
temperatures below 20~K that are larger than 1.5, ranging up to 2.5
[see, for example \citet{ris02} and references therein].  On this basis, we
believe that the combined mass of components A and B is likely to be about
15~M$_{\odot}$, although it may be as low as 6~M$_{\odot}$, or as high as
30~M$_{\odot}$.

\subsection{Molecular Gas in Relation to the YSOs}

The integrated BIMA HCO$^+(1-0)$ emission in a 1.5~km/s velocity range centered
on $-0.8$~km/s is shown in Figure~\ref{p1}d.   The bulk of the small-scale
emission detected by the BIMA interferometer is confined to a roughly circular
blob south and east of the 2MASS star.  The star itself appears to lie in a
small ``bay'' devoid of emission.  The morphology of the emission resembles
that of the $^{13}$CO~(1-0) emission mapped by \citet{fue02} around Herbig
Ae/Be stars that are just beginning to disperse their placental clouds.  

A comparison of HCO$^+(1-0)$ spectra from the BIMA interferometer data with
single-dish spectra of HCO$^+(3-2)$ in Figure~\ref{f4} illustrates the 
relative virtues of interferometric and single-dish observations.  
Figure~\ref{f4}a directly compares the
integrated emission in the blue wing where both datasets show prominent
emission.  The similarity of the maps is clear, but the much higher resolution
of the BIMA data allows us to distinguish the A and B components of the
G79.3+0.3~P1 from the larger blob of emission to the southeast. 
Figure~\ref{f4}b makes a similar comparison but with the HCO$^+(3-2)$ in the
velocity range $[-0.25,2.25]$. This range contains most of the HCO$^+(3-2)$
emission from the cloud in the single-dish data, but is  almost devoid of
HCO$^+(1-0)$ emission in the  interferometer data. 

Figures~\ref{f4}c and \ref{f4}d show spectra from the two HCO$^+$ datasets at
the points where the integrated emission peaks in the blue wing (labelled ``W''
in Figure~\ref{f4}a) and in the main part of the line (labelled ``M'' in
Figure~\ref{f4}b).  The upper spectra show the single-dish data from the JCMT,
showing the presence of significant emission in the velocity range
$[-0.25,2.25]$.  The lower spectra show the BIMA HCO$^+(1-0)$ data, and were
constructed by convolving the high-angular-resolution BIMA datacube with a
Gaussian so that the effective beam size was comparable to the JCMT beam size.
Both of these spectra exhibit the absence of detected emission in this
velocity range in the BIMA data.   Presumably, the emission is
sufficiently smooth on small angular scales  that it was resolved out by the
interferometer. 

On a larger scale, the JCMT observations show that there is significant
emission in the blue wing, i.e. the velocity range $-1.75$ to $-0.25$~km/s, 
{\it only} from the small region near G79.3+0.3~P1 that is shown in
Figure~\ref{f4}a.  The BIMA data are therefore tracing the structure within a
small parcel of gas with anomalous velocities.  

We note that the absence of a counterpart to 2MASSI~2032220+402017 in the 
2MASS J and H images suggests that it lies in or beyond the IRDC.  Moreover, if
the star were embedded in the IRDC we would expect the wind from the star to
excite red and blue wings of roughly comparable strength. Although there is
also a small amount of emission at velocities near $+2.0$~km/s that seems to be
associated with the blob,  the dominance of the blue wing indicates that the
blob is a parcel of gas being  driven into the far side of the IRDC.  The red
wing of the line in the BIMA data peaks at the same location as the blue wing,
indicating that it arises in the same parcel of gas, and that the velocity 
dispersion in this parcel of gas is appreciably larger than in the rest of the
cloud. Alternatively, the red wing might arise from an unrelated weak outflow.

There is some HCO$^+(1-0)$ emission associated with the A and B components of
the  3~mm  emission. In fact, the B component appears to sit on the edge of the
brightest part of the HCO$^+$ blob.  Even in these high-density cores,
the emission  is largely in the blue wing and not in the velocity range that
contains most of the mass in the cloud.  The presence of
emission in the blue line wing indicates that the same energy source
that has disturbed the back side of the IRDC to the southeast of G79.3+0.3~P1
is also affecting the back sides of both the A and B components. 
Unfortunately, the weakness of the emission in the A and B components precludes
the detection of extended line wings that might be taken as evidence of
collapse and/or outflow.

Follow-up BIMA observations of CS$~(5-4)$, another high-density tracer, show a
compact region of enhanced emission on  the rim of the HCO$^+$ cloud closest to
the 2MASS star (see the white contours in Figure~\ref{p1}d). The absence of 
short spacings in the CS$~(5-4)$ observations compared to the HCO$^+(1-0)$
observations make them even less sensitive to the large-scale structure of the
emission.  They have, however, higher angular resolution than the HCO$^+(1-0)$ 
observations and are useful for picking out fine-scale structure.  We
note that the B component of the 3~mm emission lies within the region of CS
emission.  

\section{Conclusions}

Our Gemini N-band observations  demonstrate that the MSX 8~\micron\ source
MSX5C G079.3398+00.3415 associated with G79.3+0.3~P1 is dominated by a single
compact object that we identify with the star 2MASSI~2032220+402017.  This
object is a luminous YSO with a strong IR excess that will likely become an
early-B star when it reaches the main sequence.  Two nearby, fainter
YSOs were discovered in our N-band images, and a cluster of faint, heavily
reddened stars are visible in the 2MASS K-band images of the region.  This
concentration of stars and YSOs indicates that the IRDC is actively forming
low-mass stars, with the 2MASS star as the most massive star that has formed
from it to date. 

There are a variety of unusual features in the region around G79.3+0.3~P1 that
suggest   2MASSI~2032220+402017 is interacting with the foreground IRDC. Most
directly, the BIMA HCO$^+(1-0)$ line has a strongly enhanced blue wing in the
velocity range $-1.5$ to 0~km/s, as well as a
slight enhancement on the red wing, in a region immediately to the southeast of
the 2MASS star.  The emission in the blue wing shows considerable structure on
small angular scales.  The unusual nature of this structure is highlighted by
the fact that the rest of the cloud shows little emission in the blue wing and
very modest small-scale structure in the velocity range $[-0.25, +2.25]$~km/s that
traces most of the mass of the cloud.  The presence of small-scale structure
close to the 2MASS star is confirmed by CS$~(5-4)$ observations with even higher
angular resolution.  

The simplest interpretation of these data is that the 2MASS star is a Herbig
Ae/Be star that is forming on the far side of the IRDC.  This star must have a
warm disk that is the source of the N-band emission, but the disk cannot
produce the observed reddening at K.  Hence the star must lie behind a lot of
extinction from the IRDC.  The strong wind that would be expected  from such a
star is probably impacting the back of the IRDC, exciting the blue wing of the
HCO$^+$ line.  Although it appears to be massive, the 2MASS star is still too
young to have disrupted the IRDC.  We speculate that in another million years
or so this region may resemble a smaller version of DR15 to the south,
appearing as an HII region partially obscured by foreground dust left over from
the IRDC.

It is, at the very least, a striking coincidence that the two  condensations of
dust and gas that comprise the G79.3+0.3~P1 SCUBA source (the A and B
components in the BIMA data) lie immediately adjacent in the plane of the sky
to the most massive YSO in the cloud, and are apparently being impacted by the
same wind that is exciting the blue wing in the HCO$^+(1-0)$ line in the rest
of the IRDC.  The combined masses of the two components are comparable to the
mass of the Herbig Ae/Be star.  These two objects are the obvious candidates
for the next generation of stars to form in the vicinity of the 2MASS star.
Further observations may reveal whether the wind from the 2MASS star has 
played a role in forming the A and B components, or is just now impacting these
two pre-existing clumps, possibly driving them into collapse.

\section{Acknowledgments}
We thank H. J. Wendker for assistance with the distance to the Cygnus Rift.
We would also like to thank the anonymous referee for a careful review which 
helped us to clarify the presentation in the paper.
The JCMT is operated by the JAC on behalf of the Particle Physics and 
Astronomy Research Council of the UK, the Netherlands Organisation
for  Scientific Research, and the National Research Council of Canada.
This paper is  based partly on observations obtained at the Gemini
Observatory,  which is  operated by AURA, Inc., under a cooperative
agreement with the NSF on behalf of  the Gemini  partnership: the NSF
(USA), PPARC (UK), the NRC (Canada),  CONICYT (Chile),  the ARC
(Australia),  CNPq (Brazil) and CONICET (Argentina).  The Gemini
observations were made with OSCIR, developed by U. Florida with 
support from NASA, and operated jointly by Gemini and the U. Florida 
Infrared Astrophysics Group.
The  Berkeley-Illinois-Maryland Association (BIMA) is a consortium
consisting of the Radio Astronomy  Laboratory at U. California
(Berkeley), the Laboratory for Astronomical Imaging at  U. Illinois
(Urbana) \& the Laboratory for Millimeter-Wave Astronomy at  U.
Maryland (College Park), which operates a millimeter-wave radio 
interferometer at Hat Creek, California. The BIMA array is operated
with support from  the National Science Foundation under grants
AST-9981308, AST-9981363, and AST-9981289. FW was supported by the 
NSF (USA) under Grant No. 96-13716.  SJC received support from NASA 
under grant NAG5-10824.
This research made use of data products from the Midcourse Space
Experiment.  Processing of the data was funded by the Ballistic
Missile Defense Organization with additional support from NASA
Office of Space Science.  This research has also made use of the
NASA/ IPAC Infrared Science Archive, which is operated by the
Jet Propulsion Laboratory, California Institute of Technology,
under contract with the National Aeronautics and Space
Administration.

\begin{deluxetable}{lllr}
\tablecaption{Properties of the Gemini N-band Sources\label{3stars}}
\tablewidth{0pt}
\tablecolumns{4}
\tablehead{
   \colhead{Designation} & 
   \colhead{RA (2000)} & 
   \colhead{Dec (2000)} & 
   \colhead{Flux Density (mJy)\tablenotemark{a}}}
\startdata
2MASSI~2032220+402017 & 20 32 21.94 & +40 20 17.5 & 1130\phantom{.0} $\pm$ 115\phantom{.3 $\pm$ 5.3} \\
northeast & 20 32 22.79 & +40 20 21.7 & 40.0 $\pm$ \phantom{11}3.5 $\pm$ 5.3 \\
southwest & 20 32 20.96 & +40 20 01.3 & 17.4 $\pm$ \phantom{11}2.5 $\pm$ 1.5 \\
\enddata
\tablenotetext{a}{Errors quoted are $\pm$ internal error $\pm$ systematic
error.}
\end{deluxetable}

\begin{deluxetable}{lrrcc}
\tablecaption{Properties of the G79.3+0.3~P1 Components\label{fluxden}}
\tablewidth{0pt}
\tablecolumns{5}
\tablehead{
   \colhead{} & 
   \colhead{RA} & 
   \colhead{Dec} & 
   \colhead{F$_{\rm 3mm}$} & 
   \colhead{F$_{0.45mm}$} \\
   \colhead{Component} &
   \colhead{(2000)} & 
   \colhead{(2000)} & 
   \colhead{(mJy)} &
   \colhead{(Jy)}
   }
\startdata
A & 20 32 21.40 & 40 20 14.0 & 7.7 & 6.9 \\
B & 20 32 22.05 & 40 20 09.9 & 9.6 & 5.7 \\
C & 20 32 22.07 & 40 20 16.5 & 7.3 & \nodata \\
\enddata
\end{deluxetable}

\begin{deluxetable}{lrrrrrr}
\tablecaption{Estimated Dust Temperatures and Total Masses\label{masstab}}
\tablewidth{0pt}
\tablecolumns{7}
\tablehead{
\colhead{} &
\multicolumn{3}{c}{G79.3+0.3~P1--A} & 
\multicolumn{3}{c}{G79.3+0.3~P1--B} \\
\colhead{}     & 
\colhead{$\beta=1.5$} & \colhead{1.75} & \colhead{2.0} & 
\colhead{$\beta=1.5$} & \colhead{1.75} & \colhead{2.0} 
}
\startdata
Temperature (K)    &     51  &    20  &   13  &   23  &   14  &  10 \\
Mass (M$_{\odot}$) &    1.6  &   5.3  & 11  &  4.6  & 10  & 18 \\
\enddata
\end{deluxetable}

\clearpage

\begin{figure}
\caption{JCMT SCUBA 850~\micron\ image of the MSX
IRDC G79.3+0.3.  The peak intensity is 1.7 Jy/beam. \label{g79}}
\end{figure}

\begin{figure}
\caption{MSX 8~\micron\ image of G79.3+0.3 covering the same area as in
Figure~\ref{g79}.  The intensities in the image have been transformed using an
inverse sinh function which results in a nearly linear scale for the faint
structures that define the IRDC but compresses the bright features in DR15
logarithmicly.  The grey contour is the 0.15~Jy/beam level from the SCUBA
850~\micron\ map and delineates the filamentary region containing high column 
densities of dust.  The white box shows the area around G79.3+0.3~P1 imaged in
Figures~\ref{p1} and \ref{f4}.  The locations of the Herbig-Haro jet and YSO mentioned in the
text are shown by the small line and circle, respectively, to the west of 
G79.3+0.3~P3. \label{8mic}}
\end{figure}

\begin{figure}
\caption{\label{p1} Four images and contour maps of the
G79.3+0.3~P1 region.   All four images cover the same $90\arcsec\times 90\arcsec$
area.   The locations of the three N-band sources are marked with asterisks.  
The middle asterisk marks the position of the brightest N-band source which is
coincident with the star 2MASSI~2032220+402017 within the pointing accuracy.
a) {\it upper left} 2MASS K$_S$ image. The field covered by the Gemini OSCIR
N-band mosaic is outlined in white.  The units for all the contours are Jy/beam.
b) {\it upper right} JCMT SCUBA 450~\micron\ contour map. 
c) {\it lower left} BIMA 3 mm continuum contour map. 
d) {\it lower right} BIMA HCO$^+$~(1-0) and CS$~(5-4)$ contour maps. The HCO$^+$
signal has been averaged over a velocity range 1.5~km/s centered on
$-$0.8~km/s; the contour levels can be converted to K~km/s by multiplying by a
factor 12.0. The CS signal has been averaged over a velocity range 1.0~km/s
centered on  +0.25~km/s; the contour levels can be converted to K~km/s in this
case by multiplying  by a factor 2.25.}
\end{figure}

\begin{figure}
\caption{\label{f4} A comparison of the BIMA HCO$^+(1-0)$ interferometer data
with the JCMT HCO$^+(3-2)$ single-dish data for the G79.3+0.3~P1 region.   The
images in \ref{f4}a and \ref{f4}b cover the same $90\arcsec\times 90\arcsec$ area
as the images in \ref{p1}.   For reference, the locations of the three  N-band
sources are marked with asterisks, and the black contours show the integrated
emission of the BIMA HCO$^+(1-0)$ line, as in figure \ref{p1}d.   
a) {\it upper left} JCMT HCO$^+(3-2)$ (white contours) in the velocity range
$[-1.75, -0.25]$ containing the blue wing of the line. The units for the white
contours are K~km/s.   The location of the spectra shown in Figure~\ref{f4}c is
indicated by the white ``W''.
b) {\it upper right}  JCMT HCO$^+(3-2)$ (white contours) in the velocity range
$[-0.25, 2.25]$ containing the bulk of the HCO$^+$ emission from the cloud.  
The units for the white contours are K~km/s.  The location of the spectra shown
in Figure~\ref{f4}d is indicated by the white ``M''.  
c) {\it lower left} Spectra of HCO$^+(3-2)$ taken with the JCMT and 
of HCO$^+(1-0)$ taken with BIMA near the peak
of emission in the blue wing (location ``W'' in \ref{f4}a).  
d) {\it lower right} Spectra of HCO$^+(3-2)$ and JCMT HCO$^+(1-0)$ near the
peak of emission in the line center (location ``M'' in \ref{f4}b).  
In \ref{f4}c and \ref{f4}d, the vertical dotted lines mark the velocity range
$[-0.25, 2.25]$ containing the bulk of the HCO$^+$ emission.}
\end{figure}
\end{document}